\documentclass{iaus}
\usepackage{graphicx}

\title[The spiral structure of our Milky Way] 
{The spiral structure of our Milky Way}

\author[L. G. Hou \& J. L. Han]   
{L. G. Hou and J. L. Han}

\affiliation{National Astronomical Observatories, Chinese Academy of
  Sciences, China \\ email: lghou@nao.cas.cn}

\pubyear{2013}
\volume{292}  
\pagerange{106-106}  
\jname{Molecular Gas, Dust, and Star Formation in Galaxies}
\editors{Tony Wong \& J\"urgen Ott, eds}
\begin{document}

\maketitle
\begin{abstract}
  The spiral structure of our Milky Way has not yet been well
  outlined. HII regions, giant molecular clouds (GMCs) and 6.7-GHz
  methanol masers are primary tracers for spiral arms. We collect and
  update the database of these tracers which has been used in
  \cite{hhs09} for mapping the spiral structure.  

\end{abstract}

\begin{figure*}[h!tb]
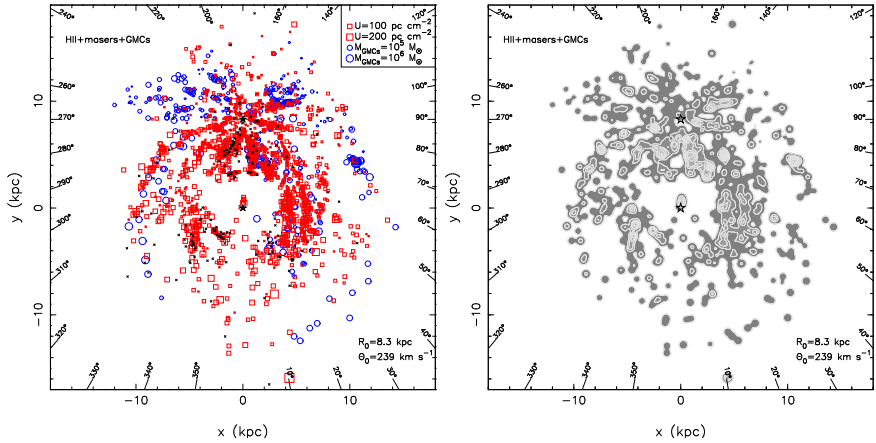

\begin{center}
\includegraphics[width=2.25in]{hou_fig1.ps} 
\includegraphics[width=2.25in]{hou_fig2.ps} 
\caption{Left: the distribution of HII regions (red squares), GMCs
  (blue circles) and 6.7-GHz methanol masers (black dots). Right: the
  contour and gray map after Gaussian extension with the amplitude of
  weighting parameter.}
   \label{fig1}
\end{center}
\end{figure*}

  The new database consists of $\sim$ 2000 HII regions, $\sim$ 1300
  GMCs and $\sim$ 800 methanol masers (6.7~GHz). If the photometric or
  trigonometric distance for any tracer is available from the
  literature, we will adopt it. Otherwise, we have to use the
  kinematic distance. We modify the V$_{LSR}$ according to the newly
  determined solar motions (~U$_0$~=~10.27~km~s$^{-1}$,
  V$_0$~=~15.32~km~s$^{-1}$ and W$_0$~=~7.74~km~s$^{-1}$,~\cite{sbd10}),
  then calculate the kinematic distances with a flat rotation curve
  (~R$_0$~=~8.3~kpc,
  $\Theta_0$~=~239~km~s$^{-1}$,~\cite{brm+11}). Very important step is that
  we weight tracers according to the excitation parameters of HII
  regions or the masses of GMCs, and a constant weight for masers. All
  three kinds of tracers are used together to outline the spiral
  structure (Fig.~\ref{fig1}). A contour and gray map is constructed
  after we made a Gaussian extension for the tracers with the
  amplitude of weighting parameter.

\end{document}